\documentclass[a4paper]{jpconfpagenum}
\usepackage{graphicx}
\usepackage{citesort}
\usepackage[numbers]{natbib}
\usepackage{color}
\usepackage{multirow}
\usepackage{hyperref}

 \hypersetup{
    pdftitle={Acoustic emission from magnetic flux tubes in the solar network},
    pdfauthor={G. Vigeesh}, 
    colorlinks=true,       
    linkcolor=black,          
    citecolor=blue,        
}
\begin{document}
\title{Acoustic emission from magnetic flux tubes in the solar network}

\author{G Vigeesh$^{1}$ and S S Hasan$^{2}$}

\address{$^{1}$ Department of Astronomy, New Mexico State University, Las Cruces, NM, U.S.A.}
\address{${^2}$ Indian Institute of Astrophysics, Koramangala, Bangalore, India}
\ead{vigeesh@nmsu.edu, hasan@iiap.res.in}

\begin{abstract}
We present the results of three-dimensional numerical simulations to investigate the excitation of
waves in the magnetic network of the Sun due to footpoint motions of a magnetic flux tube. We
consider motions that typically mimic granular buffeting and vortex flows and implement them as driving
motions at the base of the flux tube. The driving motions generates various MHD modes within
the flux tube and acoustic waves in the ambient medium. The response of the upper atmosphere to the
underlying photospheric motion and the role of the flux tube in channeling the waves is investigated.
We compute the acoustic energy flux in the various wave modes across different boundary layers defined 
by the plasma and magnetic field parameters and examine the observational implications for chromospheric and coronal heating.
\end{abstract}

\section{Introduction}
Observations of the solar surface in various filtergrams show a distribution of bright points, which are generally
associated with an underlying concentrated magnetic field located in intergranular lanes \cite{dewijn2009}.
Magnetic flux accumulates here after being swept in by granular flow forming strong vertical flux tubes at the junction
of granules. The continually changing solar photosphere perturbs these magnetic structures resulting in the generation
of magnetohydrodynamic (MHD) waves that propagate within the flux tube and also into the ambient plasma. The waves are
excited by  impulses imparted by  granules. Eventually, the flux tubes are advected towards regions of stronger
downdrafts present at the intersections of two or more granules where convectively driven vortex flows occur. The
footpoints of the magnetic field structures are dragged in the vortex resulting in, presumably, a different source of
wave excitation in the flux tube \cite{jess2009}.
The conditions that prevail in the lower atmosphere impacts the upper layers, primarily through magnetic fields. Estimating the energy supplied by various photospheric disturbance will help us to determine the fraction of the overall heating of the upper atmosphere by wave sources. In order to examine energy transport to the outer layers of the quiet solar atmosphere
mediated by magnetic fields
\cite{hasan2008,ballegooijen2011,kato2011,wedemeyer2012}, 
we need to clearly understand how the magnetic flux tube reacts to different types of perturbations.
Despite being idealized representations of the actual processes, studies have shown that various wave-driven energy transport mechanisms are possible in a magnetized atmosphere
\cite{hasan2005,khomenko2008,vigeesh2009,kitiashvili2011,fedun2011,vigeesh2012}. 
Further investigations could reveal how these processes contribute to the overall energy budget and shed light on the significance of wave related sources.
The purpose of this paper is to study two such scenarios with the aim of evaluating them in terms of their contribution to the heating of chromosphere and corona.

In this paper, we investigate the excitation of waves in magnetic flux tubes extending vertically through the solar
atmosphere. This work is an extension of our previous study
\cite{vigeesh2012}
that dealt with 3D modeling of MHD wave propagation in a magnetic flux tube embedded in a stratified atmosphere. We
showed that there are possibly more than one mechanism of wave production effective  which can lead to temporal
variations of the  emission that occur at chromospheric heights above these elements. In this paper our focus is to
investigate the generations of acoustic waves excited by these perturbations and estimate the acoustic energy flux at
different levels in the flux-tube. We calculate and compare the acoustic emission from various boundaries in the
flux-tube and also from different $\beta$ (the ratio of gas to magnetic pressure) surfaces as a result of the
excitation.

\section{Equilibrium Model and Boundary Conditions}

\begin{figure}[t]
\includegraphics[width=0.5\textwidth,height=0.45\textwidth]{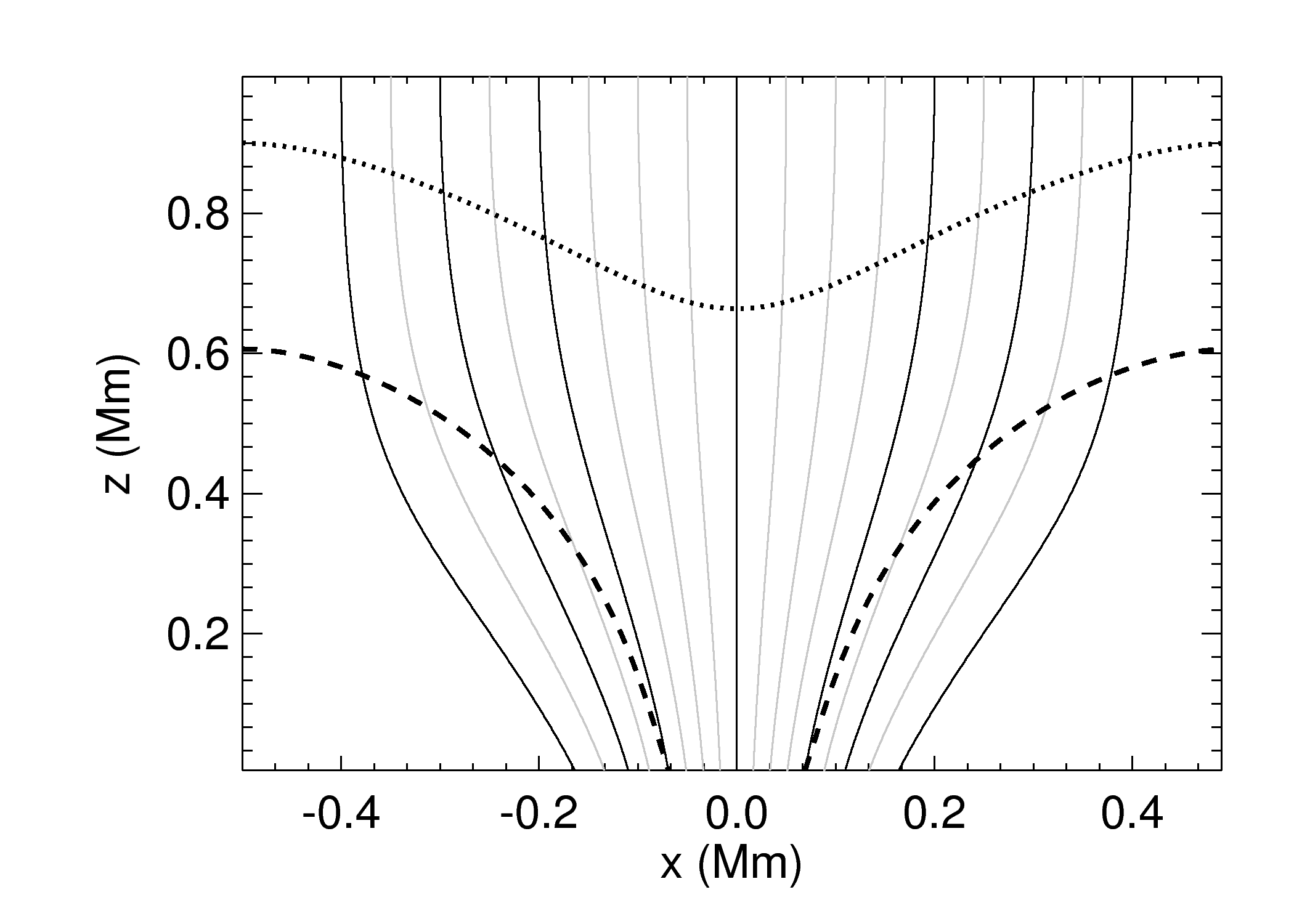}\hspace{2pc}%
\begin{minipage}[b]{14pc}\caption{\label{fig:model} Two-dimensional representation of the flux-tube model. The solid lines mark the magnetic field lines on a $x$-$z$ plane at {$y=0$~Mm}. The $\beta=1$ (\dashed) and $\beta=0.1$ (\dotted) contours are also shown.}
\end{minipage}
\end{figure}

Details of the equilibrium model that we consider in this study are given in
\cite{vigeesh2012}.  Briefly, we consider  an intense (kG field strength at the base) axially symmetric magnetic flux tube
in a stratified solar model atmosphere that is in magnetostatic equilibrium. The temperature in the model increases with 
height to include the chromospheric temperature rise. The footpoint of the flux tube is located in the solar photosphere. 
The mathematical construction of the flux tube is described in the Appendix of \cite{vigeesh2012}
and the equilibrium properties of the flux tube are given in Table~\ref{tab:equilibrium}.
\begin{table}[ht]
\caption{\label{tab:equilibrium} The equilibrium model parameters on the axis of the flux tube and ambient medium (values shown within brackets).}
\centering
\renewcommand{\arraystretch}{1.2}
\begin{tabular}{lcccccccc}
\br
Height &  T &  $\rho$ & P &  c$_{S}$ &  v$_{A}$ & B & $\beta$\\[-1ex]
&{\small (K)} & {\small(kg m$^{-3}$)} & {\small(N m$^{-2}$)} & {\small(km s$^{-1}$)} & {\small(km s$^{-1}$)} & {\small(G)} & \\
\mr
\multirow{2}{*}{$z$=1~Mm} & 7263 & 3.4 $\times$ 10$^{-8}$ & 1.6 & 8.8  & 77.7 & 161 & 64 \\[-1ex]
&{(7195)} &{(1.0 $\times$ 10$^{-7}$)} & {(4.8)} & {(8.7)} & {(39.2)} & {(141)} & {(16)}\\
\multirow{2}{*}{$z$=0~Mm} & 4768 & 1.3 $\times$ 10$^{-4}$ & 4.2 $\times$ 10$^{3}$ & 7.1 & 10.9 & 1435 & 1.9\\[-1ex]
&{(4766)} &{(4.0 $\times$ 10$^{-4}$)} & {(1.2 $\times$ 10$^{4}$)} & {(7.1)} & {(0.003)} & {(0.77)} & {(1.9 $\times$ 10$^{-7}$)}\\
\br
\end{tabular}
\end{table}
The plasma-$\beta=1$ surface essentially outlines the flux tube boundary in the lower part of the tube. In
these layers,  $\beta < 1$ inside the flux tube as shown in Fig~\ref{fig:model}.

Wave excitation is carried out by implementing velocity drivers at the bottom boundary. We use a horizontal driver to
mimic the granular buffeting motion and a torsional driver to mimic the effect of vortex-like motion. The driving motions at the bottom boundary are specified using the following velocity drivers.\\
\begin{eqnarray}
V_{x}(x,y,0,t) = \left\{ \begin{array}{l l} 
\displaystyle V_{0} \sin (2 \pi t / P) 
&\mbox{for}\quad 0 \le t \le P/2\,, \\[1ex]
\displaystyle 0
&\mbox{for}\quad 0 > t > P/2\,.
\end{array} \right. {\rm (Horizontal)}
\\
V_{\phi}(x,y,0,t) = \left\{ \begin{array}{l l} 
\displaystyle -V_{0} \tanh \left(\frac{2 \pi r}{\delta r}\right) \sin\left(\frac{2 \pi t}{P}\right)
&\quad 0 \le t \le P/2,\, \\[1ex]
\displaystyle 0
&\quad \phantom{0 >  } t > P/2.\,
\end{array} \right. {\rm (Torsional)}
\label{eq:torsional_velocity}
\end{eqnarray}

These driving motions generate slow (predominantly acoustic) and fast
(predominantly magnetic) waves in the model along with the intermediate Alfv\'{e}n wave which we do not consider in this
study.

\section{Numerical Simulation}

The three-dimensional numerical simulations were carried out using the Sheffield Advanced Code \cite{shelyag2008}.
The code uses a modified version of the set of MHD equations to deal with a strongly stratified magnetized medium. In
this study, we solve the full set of ideal magnetohydrodynamic equations in three dimensions to study the propagation of
waves in the computational domain. The domain is a 1~Mm $\times$ 1~Mm $\times$ 1~Mm cube discretized on 100 $\times$ 100 $\times$ 100 grid points. We use transmitting boundary
conditions at the top and side boundaries to allow the waves to propagate out of the simulation box.

\begin{figure}[ht]
\includegraphics[width=0.32\textwidth]{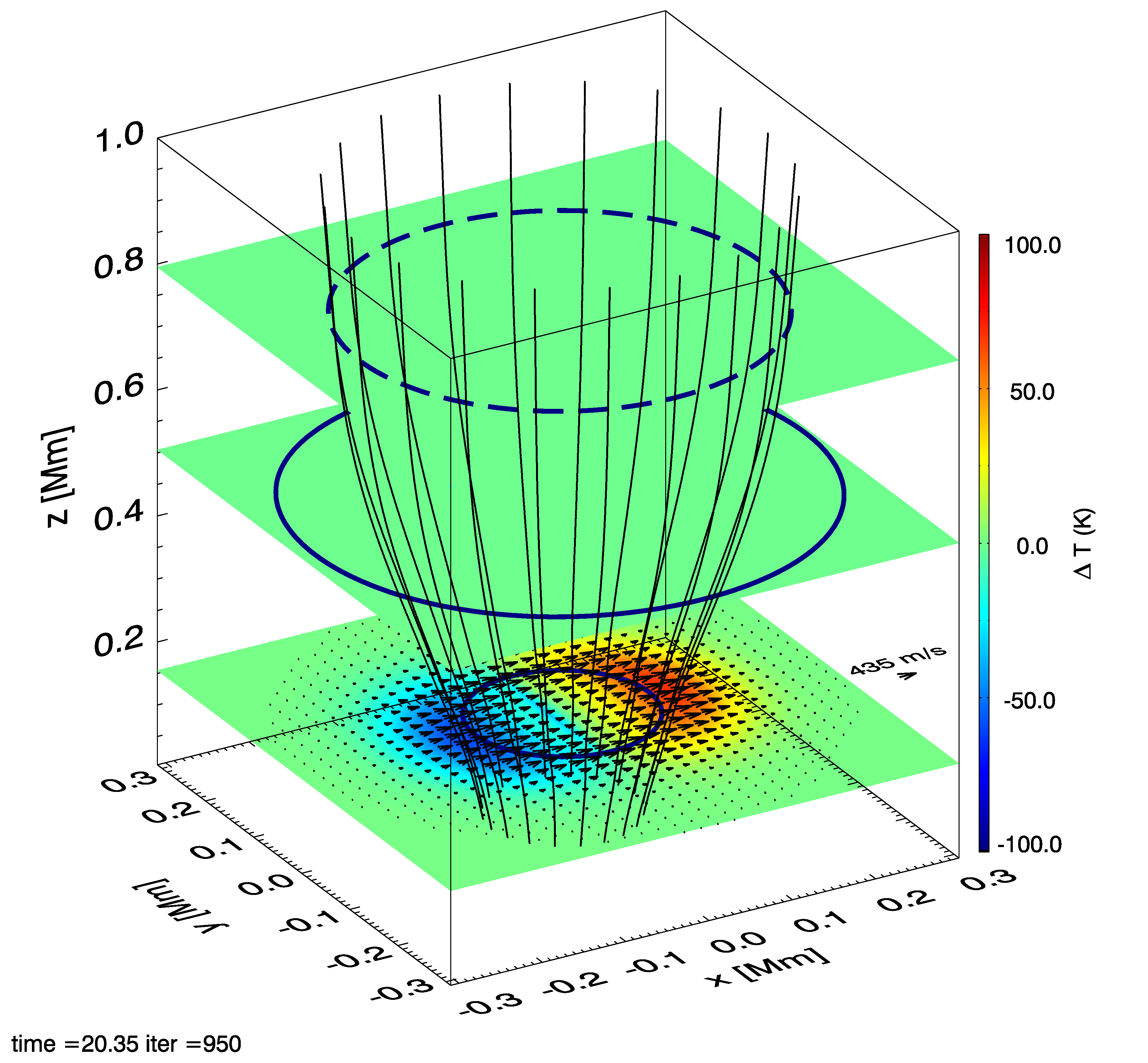}
\includegraphics[width=0.32\textwidth]{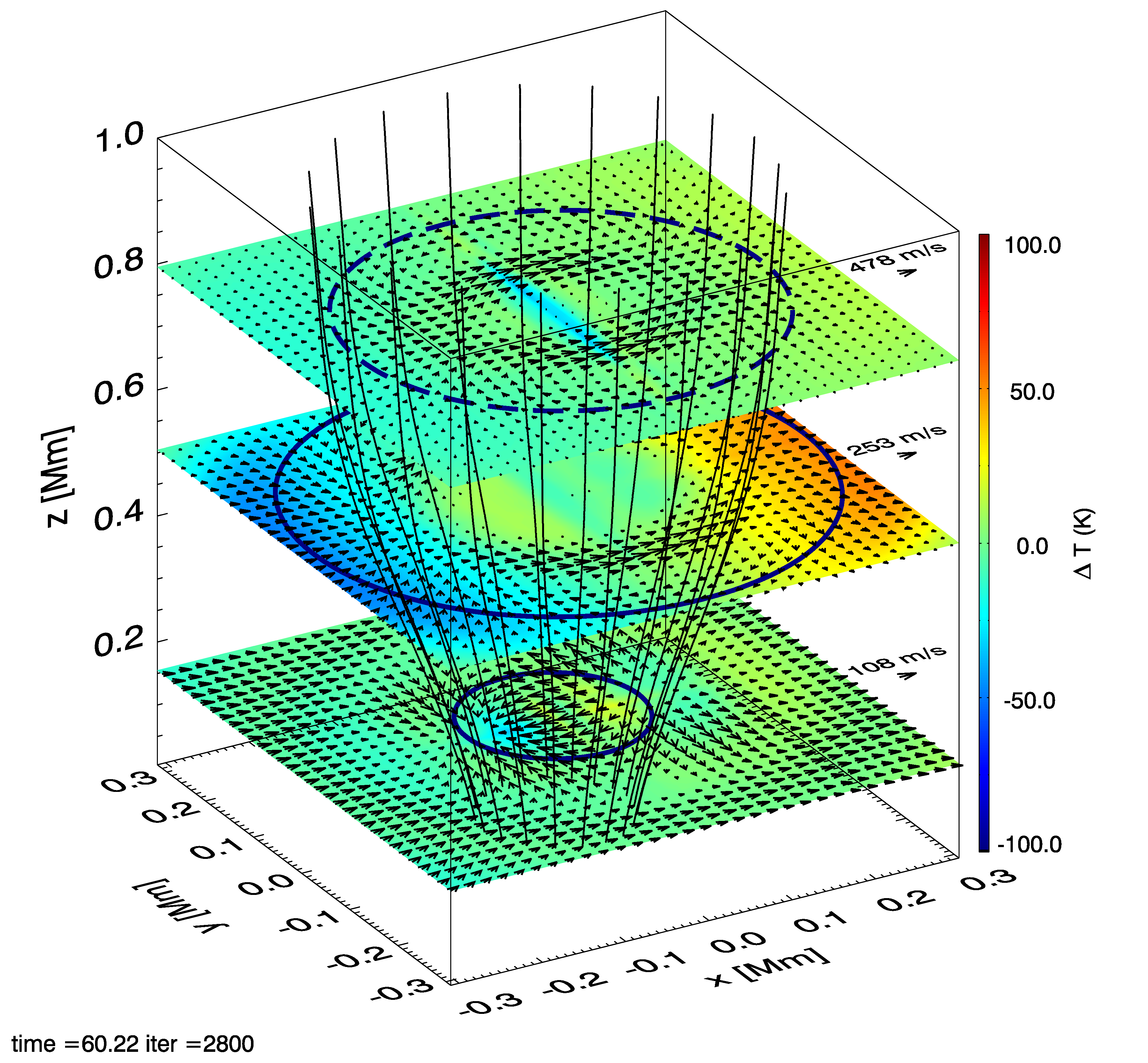}
\includegraphics[width=0.32\textwidth]{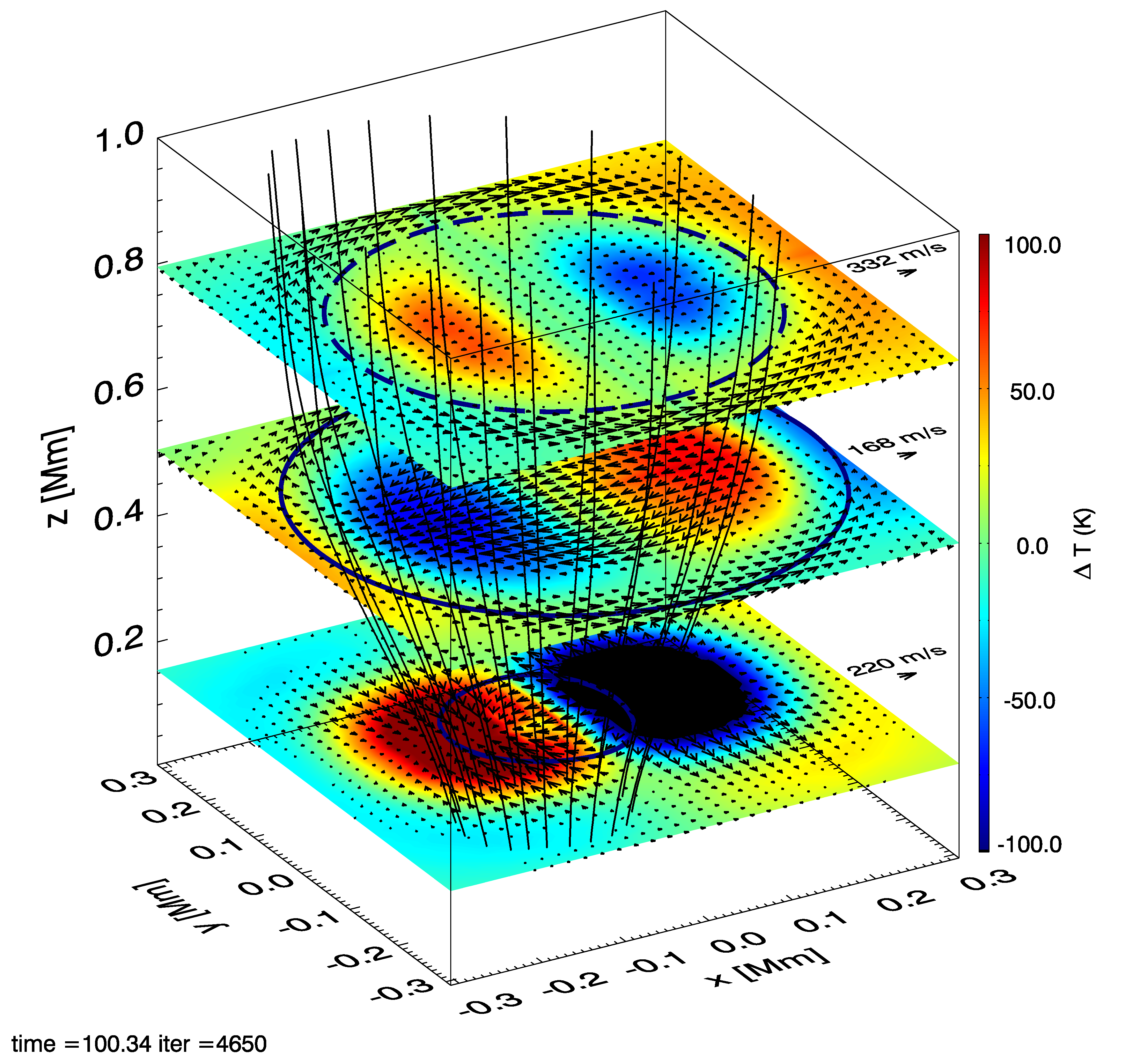}\\
\includegraphics[width=0.32\textwidth]{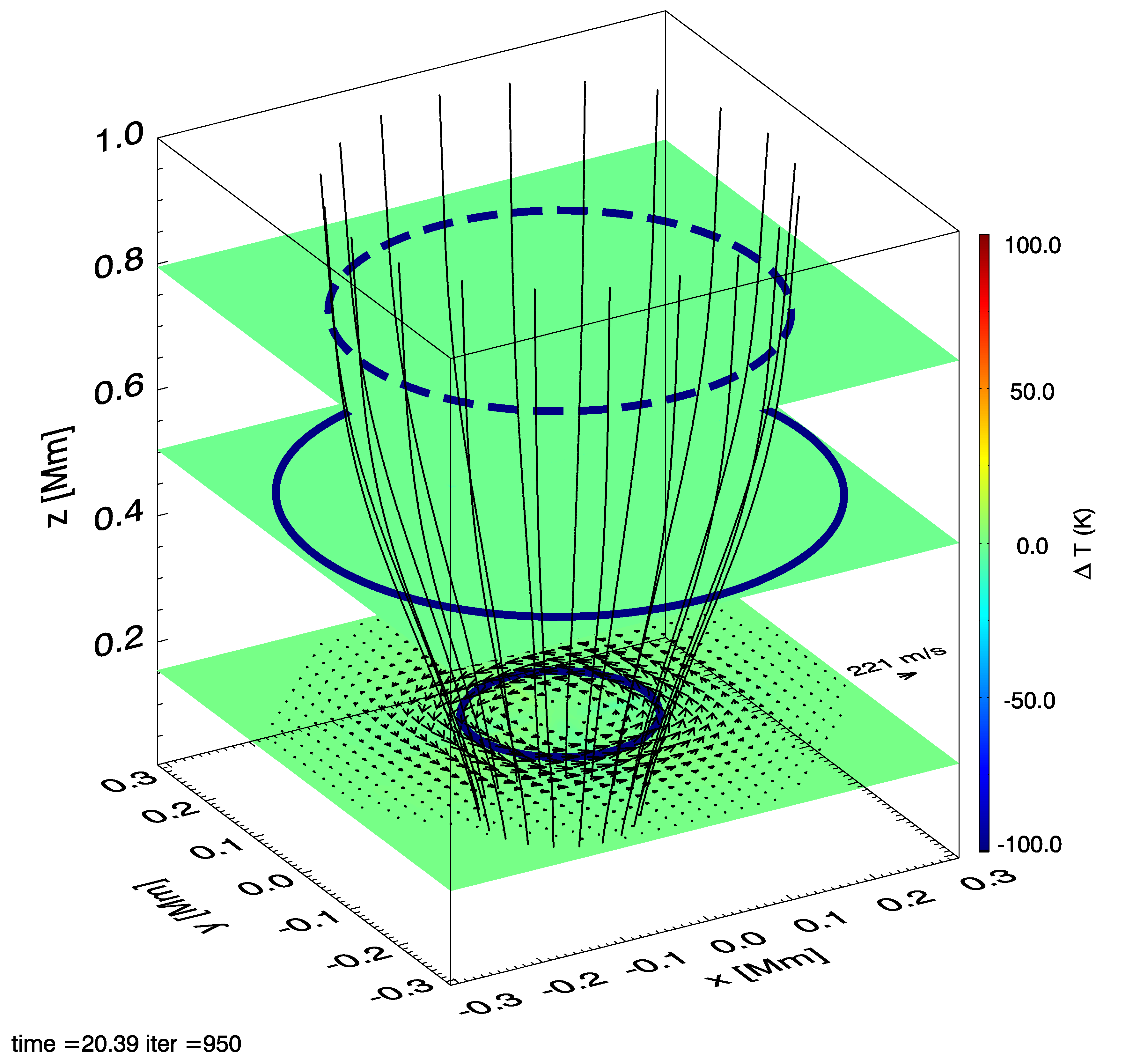}
\includegraphics[width=0.32\textwidth]{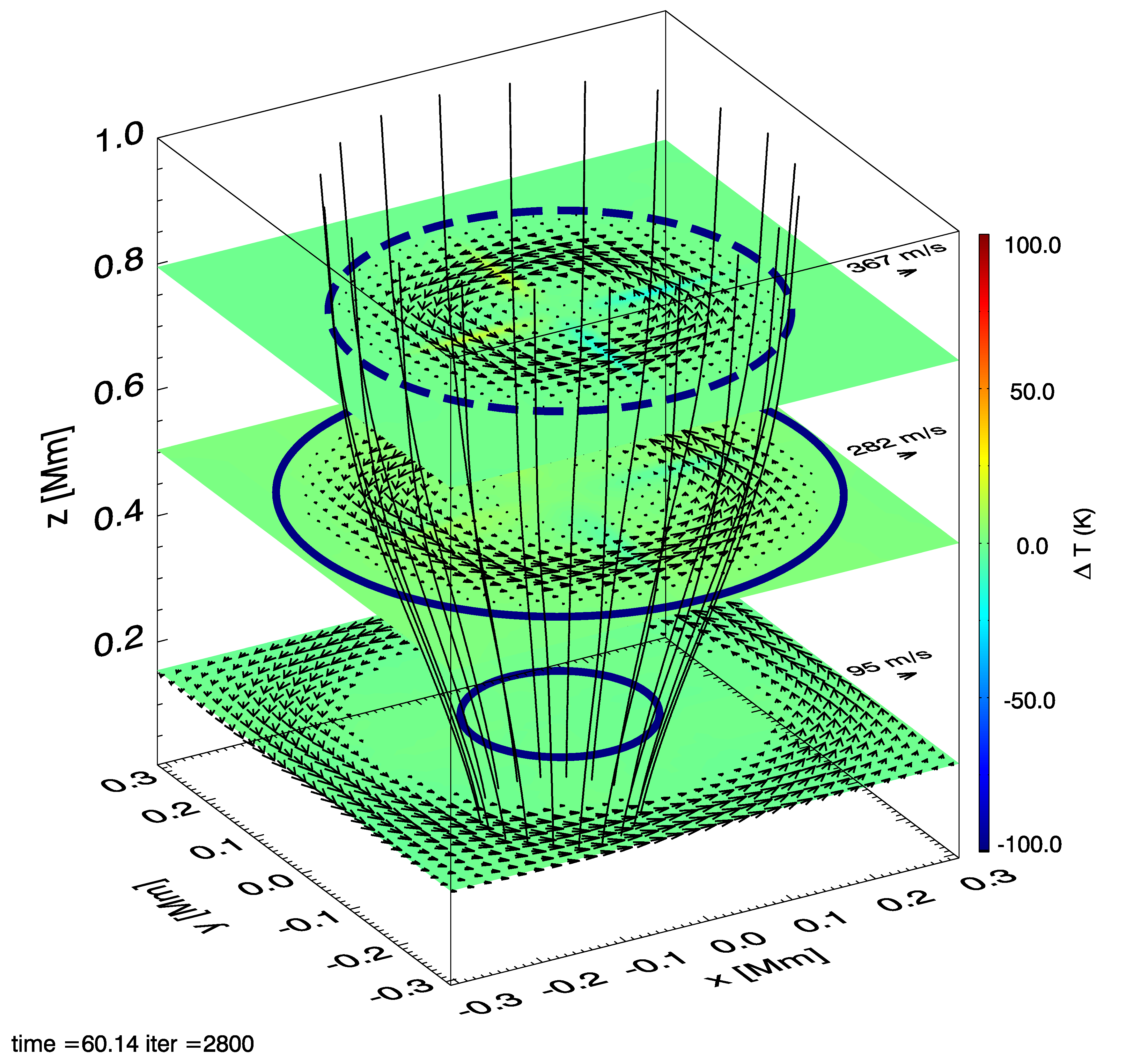}
\includegraphics[width=0.32\textwidth]{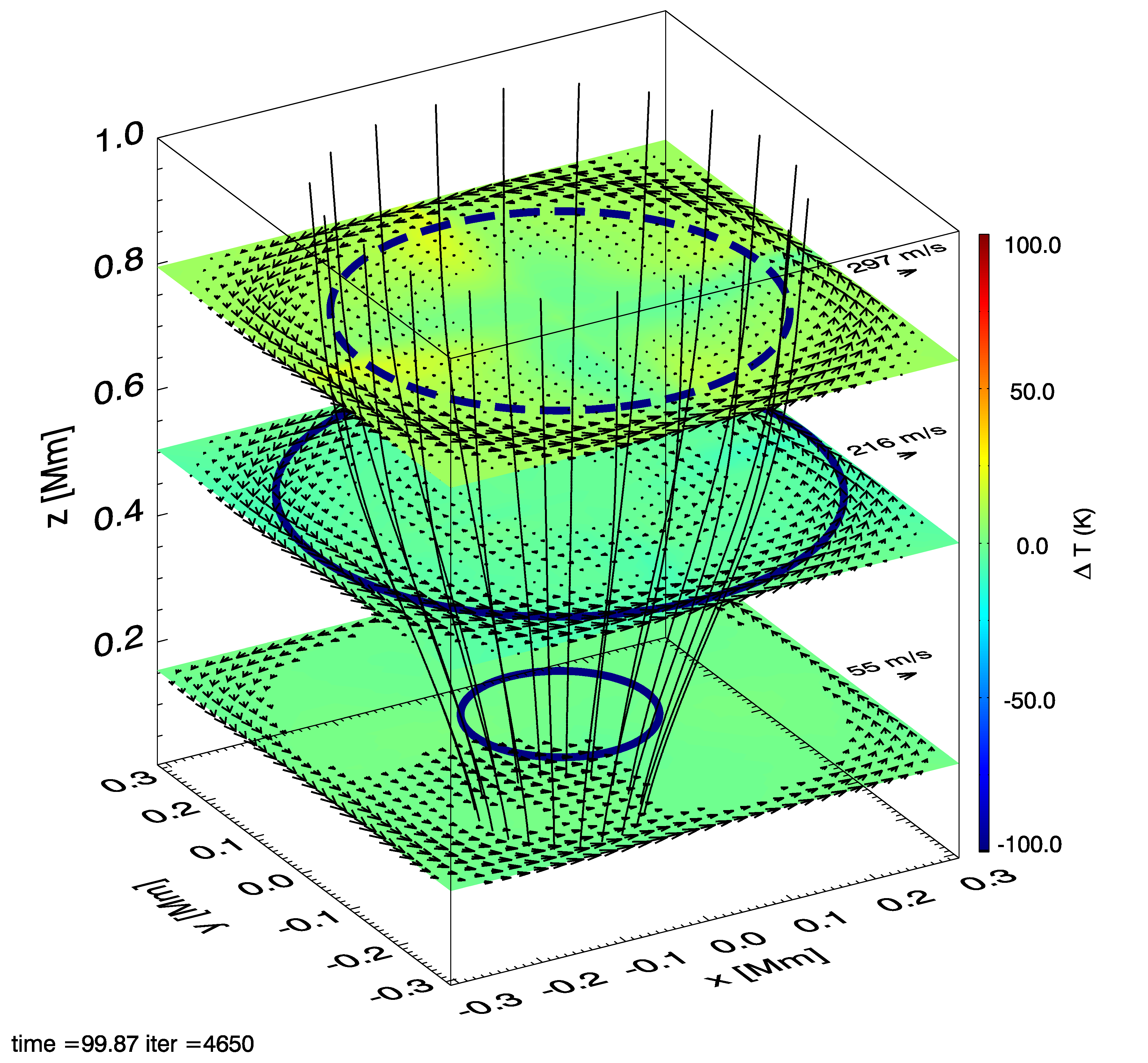}\\
\caption{\label{fig:simulation} \textit{Top:} Temperature perturbations at t=20~s,60~s, and 100~s at different heights as a result of a transversal uni-directional excitation. \textit{Bottom}: Temperature fluctuations as a result of torsional excitation. The projected velocity vectors on the $x$-$y$ plane are shown at each height. The thick blue curve depicts the
$\beta$=1 region and the blue dashed curve shows the $\beta$=0.1 region.}
\end{figure}

The simulation starts with a localized, time dependent perturbation at the bottom boundary resulting in the excitation
of various kinds of MHD modes with different strength. 
The gas pressure perturbations drive slow magneto-acoustic wave (SMAW) within the $\beta<$1 region and fast magneto-acoustic wave (FMAW) in the ambient medium where $\beta>$1. The magneto-acoustic wave propagation can be seen in the velocity
and temperature perturbation in the medium as shown in Fig~\ref{fig:simulation}. 
We notice that the horizontal excitation generates strong SMAW within the flux tube as can be seen in the temperature fluctuations. Since the SMAW propagates along field lines, we also see strong velocity parallel to the field lines. The torsional excitation on the other hand generates SMAW that
are weaker than those generated by the horizontal driver. 
In the previous study \cite{vigeesh2012}, we estimated the
energy transport by MHD waves for the two driving cases. The energy fluxes were calculated on a representative field
line and we showed that there is a strong acoustic flux associated with SMAW in the case of the horizontal excitation,
which is two orders of magnitude more than that for the case of torsional excitation. However, these studies were
restricted to a single field line and hence it was not possible to assess the response of the whole flux tube to the
different perturbations that were considered. In this paper, we extend the previous study by looking at the flux
tube as a whole and estimating the total acoustic emission from various physical surfaces relevant to the model.
Due to the nature of the drivers, the analysis on a single field line depends wholly on the location of the footpoint of the chosen field line with respect to the driving motion. Especially, in the case of a horizontal motion, where the strongest perturbations is localized to a small region on either side of the $y=0$ plane.  
Also, since the velocity driver acts in region where $\beta <1$ (close to the axis) as well as $\beta > 1$ (outer regions), a field line located in either of these region sees a different mode of magneto-acoustic wave. Considering an ensemble of field lines will give a better idea about the reaction of the flux tube as a whole to the excitation at the base.

\section{Results}
\subsection{Longitudinal \& Lateral emission}
\begin{figure}[t]
\includegraphics[width=0.5\textwidth]{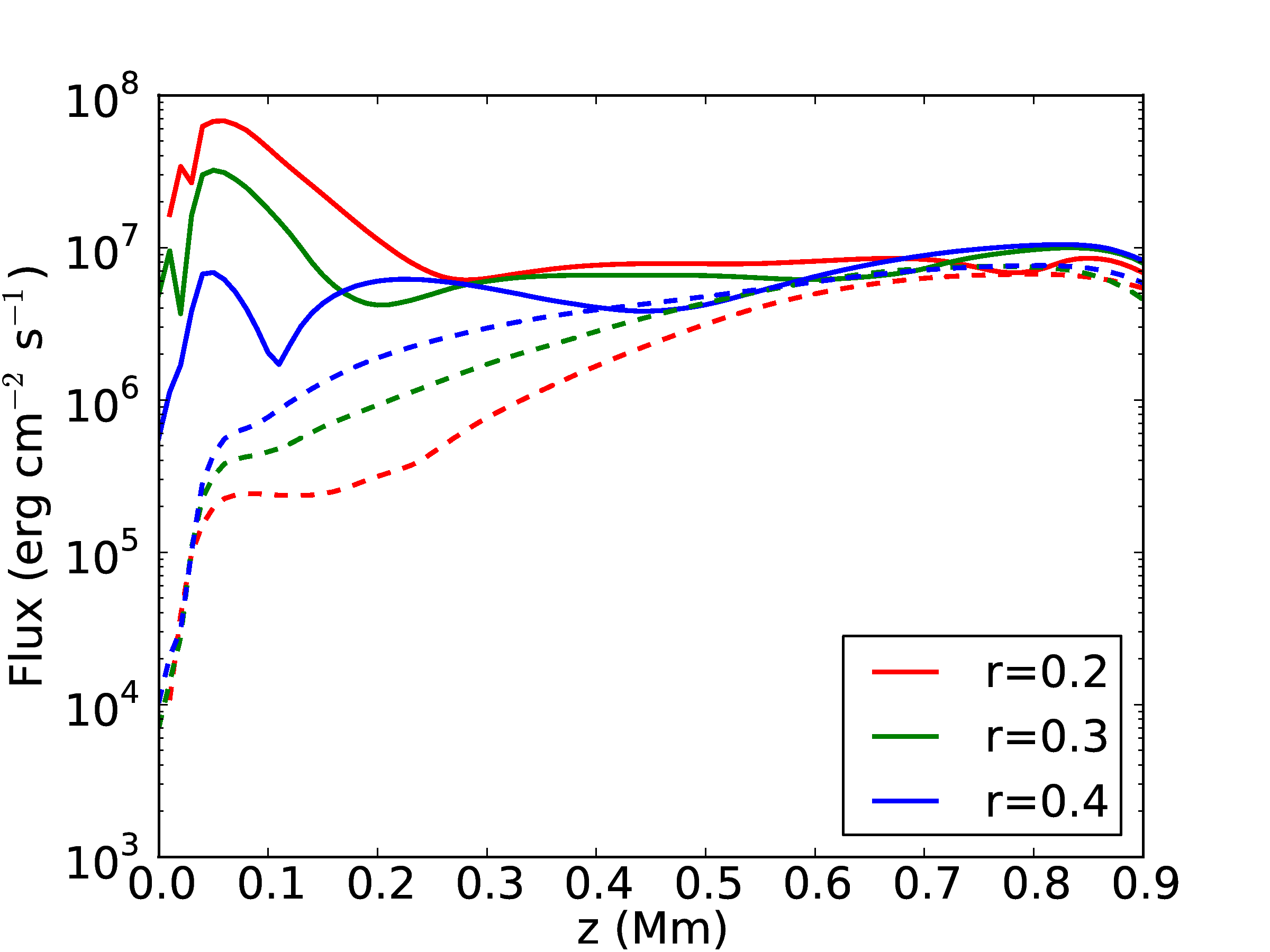}\includegraphics[width=0.5\textwidth]{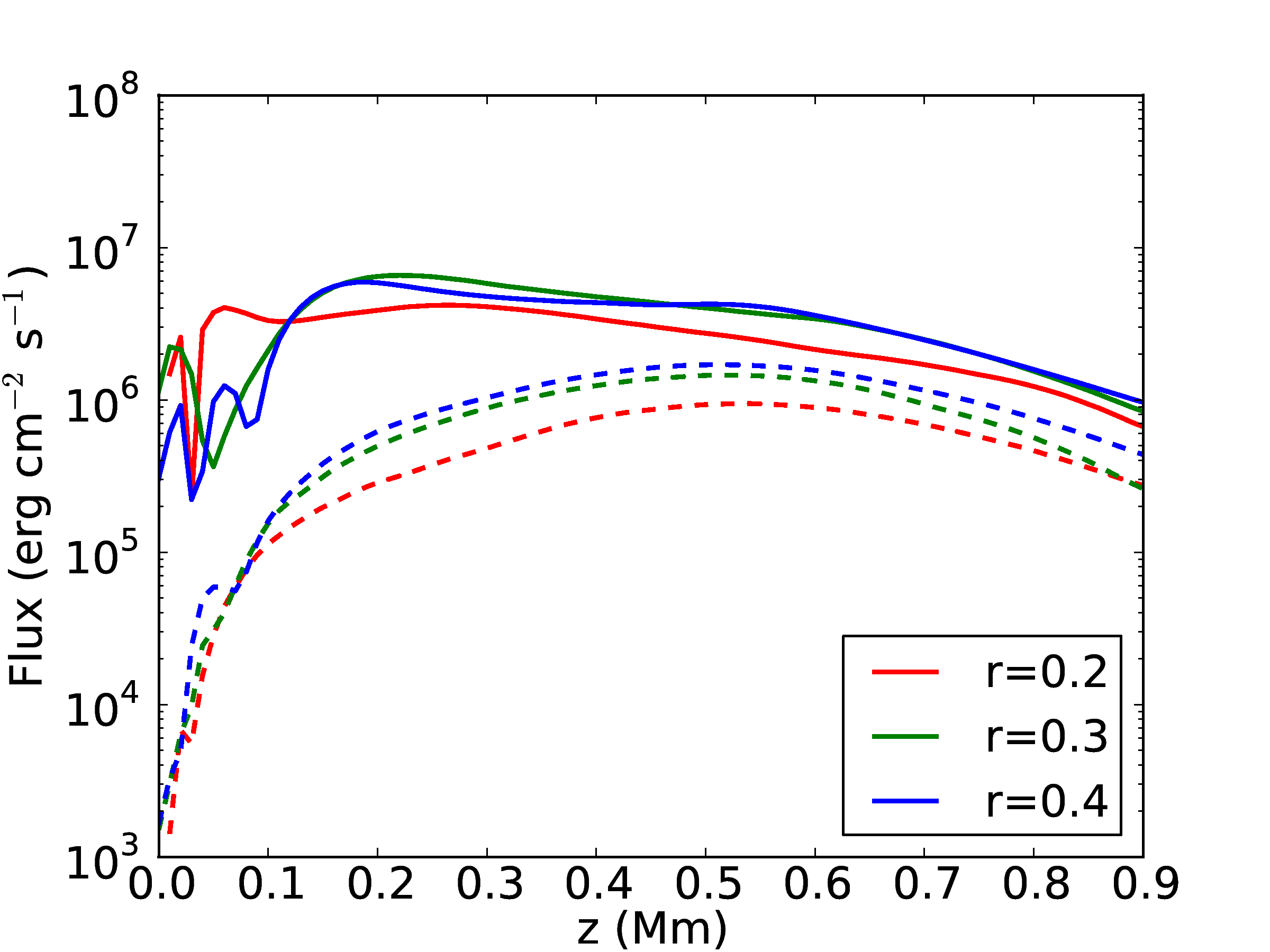}
\caption{\label{fig:latflux}\textit{Left}: Time-averaged longitudinal acoustic flux on the three isosurfaces for the horizontal (Solid lines) and 
torsional(dashed lines) excitation cases. \textit{Right} Normal acoustic flux on the three isosurfaces.}
\end{figure}

Apart from acting as a conduit for MHD waves to the overlying layers, the perturbed flux tube also transfers energy to
the ambient medium in the form of acoustic waves. It is interesting to look at acoustic energy flux leaking out of the
flux tube boundary and how it depends on the nature of the excitation that the flux tube undergoes. Since we consider
here a thick flux tube, a strict definition of the flux tube boundary is ambiguous. However, for the purpose of this
study, we choose the surfaces of equal magnetic potential or the magnetic isosurface as representative levels in the 
flux tube for purposes of calculating the fluxes. The cross-section of the magnetic isosurface at any given height is a
circle centered at the axis, due to the cylindrical symmetry of the initial model. This allows us to define a circle at
a radial distance $r$ of the flux tube on any horizontal plane. 
We calculate the acoustic flux on three magnetic
isosurfaces crossing radial distance of $r=0.2~$Mm, $0.3~$Mm and $0.4~$Mm from the axis of the flux tube at $z=1~$Mm. The magnetic field lines that define this isosurface can be identified from the equilibrium model and they remain
the same throughout the simulation, allowing us to calculate physical quantities on this surface as it evolves. The
velocity vector at any point on this surface can be decomposed into three orthogonal components, \textit{viz.}~parallel
(${v_{s}}$), normal (${v_{n}}$) and azimuthal (${v_{\phi}}$) component. The calculation of these component for a single
field line is described in \cite{vigeesh2012}.
Using these velocity components for a collection of field lines defining various isosurfaces, we calculate the parallel and normal acoustic fluxes according to,
\begin{equation}
F_{\rm s,n} = \Delta p {v_{\rm s,n}}
\end{equation}
where $\Delta p$ is the gas pressure perturbation from the equilibrium.  Acoustic fluxes are calculated on field lines corresponding to three magnetic isosurfaces. In Figure~\ref{fig:latflux}, the left
panel shows the time averaged parallel acoustic fluxes ($F_{\rm s}$) on field lines associated with different
isosurfaces as a function of height.
The horizontal excitation results in relatively stronger acoustic emission compared to torsional excitation in the lower part of the flux tube. But, the averaged fluxes tend to be similar as we go higher up in the atmosphere, since in the horizontal excitation case, only a small fraction of field lines on either side of the tube partake in transporting the longitudinal fluxes. The right
panel of Fig.~\ref{fig:latflux} shows the time averaged acoustic flux directed normal to the flux-tube boundaries, which is
significantly lower than the fluxes directed along the field lines in the tube. 

\subsection{$\beta$-surface emission}
The surface of equipartition ($\beta$=1 layer) between thermal and magnetic energy density influences the MHD modes by
acting as region of wave transmission and conversion. As a preliminary step towards understanding this region more
closely, we look at the acoustic energy flux crossing normal to the surfaces of constant plasma $\beta$ surface ($\beta$
isosurfaces). In this study, we mainly focus on the $\beta=1$  isosurface, since this is where the different MHD 
modes undergo strong coupling which eventually leads to the generation of Alfv\'{e}n waves in the upper chromosphere. The whole process proceeds in two stages. Initially, the magneto-acoustic modes encounter the equipartition zone as they propagate to the upper layers of the atmosphere. The energy in the modes get redistributed partially by transmission and partially by conversion to other modes. The transmission and conversion depends on the properties of the wave and the background magnetic field across this zone of influence
\cite{cally2007}.
In the second stage, the mode converted magnetic wave (FMAW) propagates up to a certain height after which it gets partially reflected down and partially gets converted to Alfv\'{e}n waves due to steep gradients in the Alfv\'{e}n speed
\cite{khomenko2012}.
A better understanding of the energy distribution during the initial stage where mode coupling occurs across the $\beta=1$ surface is nessasary to further evaluate the energy available for Alfv\'{e}n wave. To this end, we look at the acoustic flux directed normal to the surface of $\beta=1$ and compare it with that from
a $\beta=0.1$ isosorface which lies well above in the atmosphere. It should be noted that in the lower part of the flux tube, the $\beta=1$ surface normal points towards the axis of the flux-tube.

\begin{figure}[t]
\includegraphics[width=0.5\textwidth]{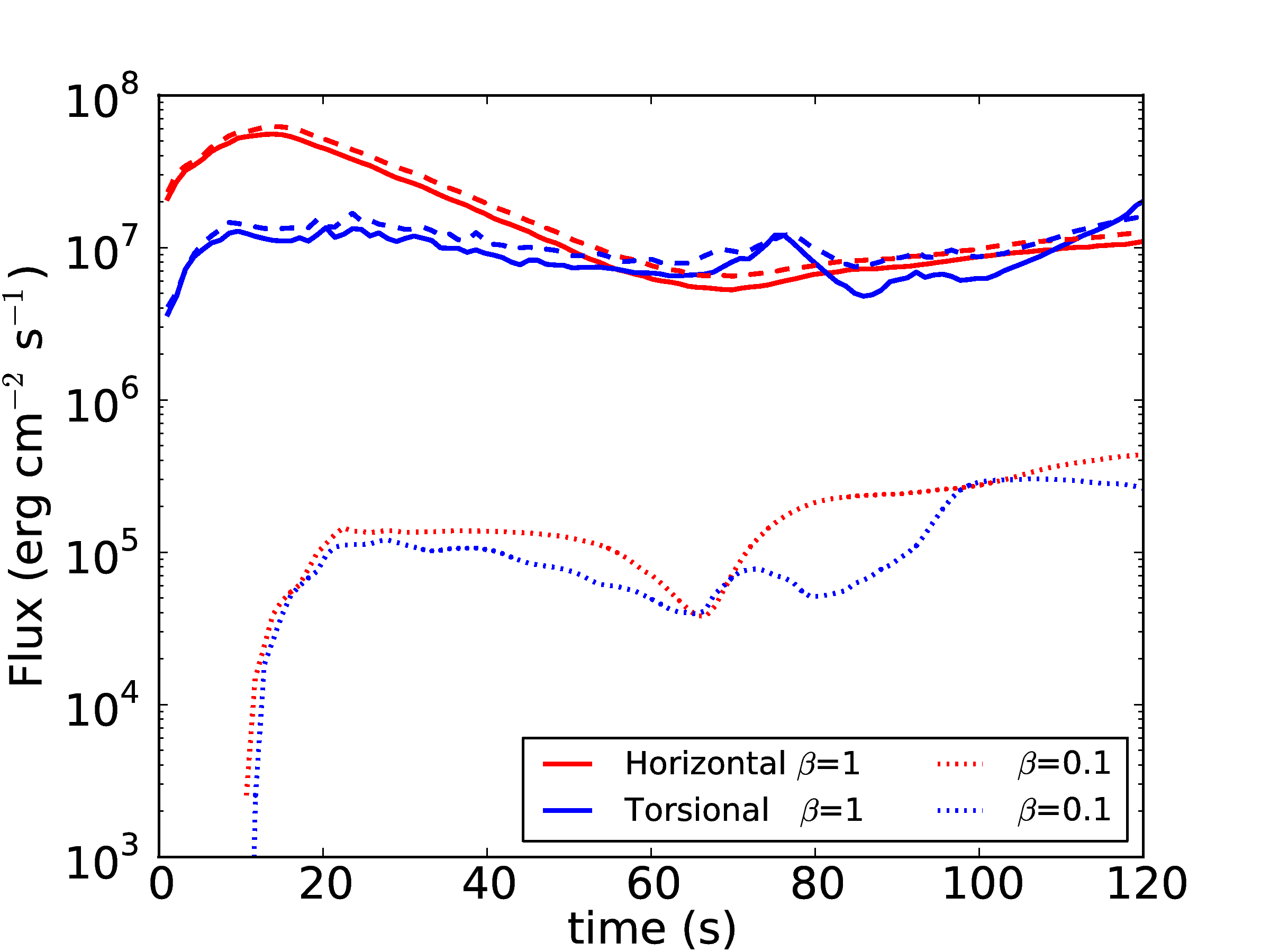}\hspace{2pc}%
\begin{minipage}[b]{14pc}\caption{\label{fig:betaflux}Time evolution of the normal acoustic flux on $\beta$=1 (solid line) and 
$\beta$=0.1 (dotted) surface for horizontal (red line) and torsional (blue) excitation of a magnetic flux tube}
\end{minipage}

\end{figure}

Figure~\ref{fig:betaflux} shows the spatially averaged acoustic flux directed normal to the surfaces of constant $\beta$ for $\beta$=1
and 0.1 as a function of time. We clearly see that the acoustic emission from the $\beta$=1 surface is stronger compared
to the emission on the $\beta$=0.1 surface for both horizontal and torsional excitations. The $\beta$=1 surface responds
more efficiently to a horizontal uni-directional motion at the foot-point by emitting more acoustic flux normal to the
surface. Our analysis is limited by the fact that we consider a strong flux tube where the $\beta=1$ surface dips below the bottom boundary near the axis of the flux tube. To have a better understanding about this layer, we need to look at flux tubes models with $\beta=1$ surfaces at different levels in the atmosphere.

\section{Summary \& Conclusions}
Three dimensional numerical simulation of wave propagation in a magnetic flux tube embedded in a solar atmosphere were
carried out.
We investigated two types of excitation mechanism, viz. transversal and torsional, that are characteristic processes by
which waves can be generated in the real solar atmosphere. We calculated the acoustic energy components at various
levels in the flux tube as well as on the surfaces defined by constant plasma $\beta$ of the flux tube due to the driving 
footpoint motions.
We observe that the acoustic power is predominantly directed vertically upward along the flux tube in both cases. The
lateral acoustic emission from the boundary of the flux tube in both cases of horizontal and torsional excitations is 
much lower. 
The magnetic flux tube acts as an efficient conduit for acoustic waves, with negligible acoustic leakage from the tube boundary. Most of the acoustic energy produced due to photospheric disturbances are efficiently transported and are made available to higher layers regardless of the source of the disturbance. As far as the comparison between the two excitation scenarios in terms of acoustic energy transport to upper layers is concerned, we have not been able to conclusively point out which of the two mechanisms is more efficient. This is partly due to the axisymmetric equilibrium model that we have used and the idealistic driving mechanisms that we consider in the study. Nevertheless, our analysis supports that the magnetic field mediates the coupling between various photospheric disturbances and the upper layers. However meagre their contribution to the overall energy output is, these scenarios must be considered when evaluating the available energy sources for the heating of chromosphere and corona.
The surface of equipartition between magnetic and
thermal energy density ($\beta=1$) is a strong source of acoustic emission in both excitation scenarios and would be the
focus of future investigations.
\pagebreak
\bibliography{ms_vigeesh}
\end{document}